\documentstyle[aps,12pt]{revtex}

\begin{document}

\title{Closed System of Equations on a Brane}

\author{Yu.~V.~Shtanov\footnote{E-mail: shtanov@bitp.kiev.ua}}
\address{Bogolyubov Institute for Theoretical Physics, Kiev 03143, Ukraine}

\date{\today}

\maketitle

\begin{abstract}
We obtain a generic closed system of equations on a brane that describes its
inner evolution and give a method for extending solutions on the brane to the
bulk. We also discuss the cosmological implications of the closed system of
equations obtained.  We consider bulk spaces with both spacelike and timelike
extra dimension, with and without the $Z_2$ symmetry of reflection relative to
the brane.
\end{abstract}

\pacs{PACS number(s): 04.50.+h, 98.80.Hw}

The idea that our four-dimensional world can be described as a timelike
hypersurface (brane) embedded in or bounding a five-dimensional manifold has
been extensively studied already for several years.  In its ideal formulation,
it is only the evolution on the brane that is observable by experimenters
themselves residing in the four-dimensional braneworld, and the role of the
five-dimensional bulk is to provide a special form of the laws that govern this
evolution on the brane, for example, the unusual form of the law of gravity on
relatively small scales, as was observed in the seminal papers \cite{RS}.

In solving equations for the brane, there naturally arises the question about
the additional conditions that must be satisfied by the solution in the bulk.
Usually, one considers another brane or system of branes on which one can
impose certain boundary conditions and require that the solution be nonsingular
in the space between the branes. Very often, however, no such additional
conditions in the bulk are actually imposed while solving field equations in
the neighborhood of the brane of interest. Such an approach has some right to
exist since, as was pointed out, the only observable evolution is the evolution
on a single braneworld.  In this letter, we would like to point out that, {\em
in this approach\/}, and in the absence of matter fields in the bulk, all the
information about the evolution on the brane is contained in a system of closed
equations [Eqs.~(\ref{constraint-new})--(\ref{s-new}) below], without the
necessity of solving field equations in the bulk.  A similar observation in the
cosmological context was previously made in \cite{BDEL} and recently noted in
\cite{KS} for the linearized equations in the case of $Z_2$ symmetry.  In
\cite{M}, it was argued that cosmological density and curvature perturbations
on sufficiently large spatial scales are described by a closed system of
equations on the brane. Here, we would like to point out that this feature is
rather general.

We also describe a method for constructing the extension of solution on the
brane to the bulk using Gaussian normal coordinates in the neighbourhood of the
brane.

In most of the braneworld models, the extra dimension is assumed to be
spacelike, so that the brane is embedded in a Lorentzian five-dimensional
manifold. However, no physical principle appears to prevent us from considering
a complementary model in which the extra dimension is timelike, so that the
brane is a Lorentzian boundary of a five-dimensional space with signature
$(-,-,+,+,+)$.  Such general embeddings in the case of $Z_2$ symmetry were
considered in \cite{Kofinas} and also in \cite{SS}.  Models with both
signatures of the five-dimensional bulk space will be examined in this paper.

To begin with, we consider a particular case where a four-dimensional
hypersurface (brane) ${\mathcal B}$ is the boundary of a five-dimensional
Riemannian manifold (bulk) ${\mathcal M}$ with nondegenerate Lorentzian induced
metric. The action of the theory has the natural general form
\begin{equation} \label{action}
S = M^3  \left[ \int_{\mathcal M} \left({\mathcal R} - 2 \Lambda \right) - 2
\epsilon \int_{{\mathcal B}} K  \right] + \int_{\mathcal M} L_5 (g_{ab}, \Phi)
+ \int_{{\mathcal B}} \left( m^2 R - 2 \sigma \right) + \int_{{\mathcal B}} L_4
(h_{ab}, \phi) \, .
\end{equation}
Here, ${\mathcal R}$ is the scalar curvature of the five-dimensional metric
$g_{ab}$ on ${\mathcal M}$, and $R$ is the scalar curvature of the induced
metric $h_{ab} = g_{ab} - n_a n_b$ on ${\mathcal B}$, where $n^a$ is the vector
field of the {\em inner\/} unit normal to ${\mathcal B}$. The quantity $K =
K_{ab} h^{ab}$ is the trace of the symmetric tensor of extrinsic curvature
$K_{ab} = h^c{}_a \nabla_c n_b$ of ${\mathcal B}$ in ${\mathcal M}$. The
parameter $\epsilon = 1$ if the signature of the bulk space is Lorentzian, so
that the extra dimension is spacelike, and $\epsilon = -1$ if the signature is
$(-,-,+,+,+)$, so that the extra dimension is timelike. The symbols $L_5
(g_{ab}, \Phi)$ and $L_4 (h_{ab}, \phi)$ denote, respectively, the Lagrangian
densities of the five-dimensional matter fields $\Phi$ and four-dimensional
matter fields $\phi$ the dynamics of which is restricted to the brane
${\mathcal B}$ so that they interact only with the induced metric $h_{ab}$.
Some of the fields $\phi$, in principle, may represent restrictions of some of
the fields $\Phi$ to the boundary ${\mathcal B}$. All integrations over
${\mathcal M}$ and over ${\mathcal B}$ are taken with the natural volume
elements $\sqrt{- \epsilon g}\, d^5 x$ and $\sqrt{- h}\, d^4 x$, respectively,
where $g$ and $h$ are the determinants of the matrices of components of the
metric on ${\mathcal M}$ and of the induced metric on ${\mathcal B}$,
respectively, in coordinate basis. The symbols $M$ and $m$ denote the Planck
masses of the corresponding spaces, $\Lambda$ is the five-dimensional
cosmological constant, and $\sigma$ is the brane tension.

In this paper, we systematically use the notation and conventions of
\cite{Wald}. In particular, we use the one-to-one correspondence between
tensors in ${\mathcal B}$ and tensors in ${\mathcal M}$ which are invariant
under projection to the tangent space to ${\mathcal B}$, i.e., tensors
$T^{a_1 \cdots a_k}{}_{b_1 \cdots b_l}$ such that
\begin{equation}
T^{a_1 \cdots a_k}{}_{b_1 \cdots b_l} = h^{a_1}{}_{c_1} \cdots
h^{a_k}{}_{c_k} h_{b_1}{}^{d_1} \cdots h_{b_l}{}^{d_l} T^{c_1 \cdots
c_k}{}_{d_1 \cdots d_l} \, .
\end{equation}

The term containing $m^2 R$ is sometimes missing from the action, or its
contribution is missing from the equations of motion. However, in general,
this term is essential since it is generated as a quantum correction to the
matter action in (\ref{action}); in this paper, we take it into account
following \cite{CH,DGP,Shtanov}.

Variation of action (\ref{action}) gives the equation of motion in the
five-dimensional bulk:
\begin{equation} \label{bulk}
{\mathcal G}_{ab} + \Lambda g_{ab} = {1 \over M^3} T_{ab} \, ,
\end{equation}
and on the brane ${\mathcal B}$:
\begin{equation} \label{brane}
m^2 G_{ab} + \sigma h_{ab} = \epsilon M^3 S_{ab} + \tau_{ab} \, ,
\end{equation}
where ${\mathcal G}_{ab}$ and $G_{ab}$ are the Einstein's tensors of the
corresponding spaces, $S_{ab} \equiv K_{ab} - K h_{ab}$, and $T_{ab}$ and
$\tau_{ab}$ denote the five-dimensional and four-dimensional stress--energy
tensors of matter, respectively.  It is the presence of the tensor $S_{ab}$ in
the equation of motion (\ref{brane}) that makes the inner dynamics on the brane
${\mathcal B}$ unusual.\footnote{Those interested in the derivation of
(\ref{brane}) may look into the appendix of \cite{Shtanov}.}

The Codazzi relation
\begin{equation} \label{conserve-s}
D_a S^a{}_b = {\mathcal R}_{cd} n^d h^c{}_b \, ,
\end{equation}
where $D_a$ is the (unique) derivative on the brane ${\mathcal B}$
associated with the induced metric $h_{ab}$, together with Eq.~(\ref{brane})
and bulk equation (\ref{bulk}) imply the relation
\begin{equation} \label{conserve}
D_a \tau^a{}_b = - \epsilon T_{cd} n^d h^c{}_b \, .
\end{equation}
Thus, the four-dimensional stress--energy tensor is covariantly conserved on
the brane if and only if the right-hand side of (\ref{conserve}) is vanishing
at the brane, in particular, if the five-dimensional stress--energy tensor
$T_{ab}$ is a linear combination of $g_{ab}$ and $h_{ab}$ at the brane. In this
paper, we assume this to be the case by setting $T_{ab} = 0$, i.e., by
considering only vacuum Einstein equation in the bulk.

Equation (\ref{brane}) involves the tensor $S_{ab}$ which is constructed from
the tensor of extrinsic curvature of the brane, so this equation is not closed
with respect to the intrinsic evolution on the brane. However, it is possible
to obtain an equation which involves only four-dimensional fields on the brane.
In doing this, we follow the procedure of \cite{BDL,SMS}.  We use the Gauss
relation on the brane:
\begin{equation} \label{gauss}
R_{abc}{}^d = h_a{}^f h_b{}^g h_c{}^k h^d{}_j\, {\mathcal R}_{fgk}{}^j +
\epsilon \left( K_{ac} K_b{}^d - K_{bc} K_a{}^d \right) \, .
\end{equation}
Contracting this relation and taking into account Eq.~(\ref{bulk}) with $T_{ab}
= 0$, one obtains the equation
\begin{equation} \label{constraint}
\epsilon \left( R - 2 \Lambda \right) + K_{ab} K^{ab} - K^2 \ \equiv \ \epsilon
\left( R - 2 \Lambda \right) + S_{ab} S^{ab} - \frac13 S^2 \ = \ 0 \, ,
\end{equation}
which we also expressed in terms of $S_{ab} = K_{ab} - h_{ab} K$ and $S =
h^{ab} S_{ab}$.  This is the well-known constraint equation on the brane from
the viewpoint of the gravitational dynamics in the five-dimensional bulk. Now
we express the tensor $S_{ab}$ in terms of the four-dimensional quantities
using Eq.~(\ref{brane}) and substitute it into Eq.~(\ref{constraint}). We
obtain
\begin{equation} \label{s}
S_{ab} = {\epsilon \over M^3} \left( m^2 G_{ab} + \sigma h_{ab} - \tau_{ab}
\right) \, ,
\end{equation}
\begin{equation} \label{closed} 
\epsilon M^6 \left( R - 2 \Lambda \right) + \left( m^2 G_{ab} + \sigma h_{ab} -
\tau_{ab} \right) \left( m^2 G^{ab} + \sigma h^{ab} - \tau^{ab}
\right) 
- {1 \over 3} \left( m^2 R - 4 \sigma + \tau \right)^2 = 0\, ,
\end{equation}
where $\tau = h^{ab} \tau_{ab}$.

Together with the equations for other fields on the brane, Eq.~(\ref{closed})
forms a system of closed equations on the brane to be solved.  Note that the
five-dimensional part of the curvature tensor in the bulk (the projected Weyl
tensor) drops from this equation because the complete trace of the Gauss
relation (\ref{gauss}) has been taken. Once its solution is obtained, one can
calculate the tensor of extrinsic curvature using Eq.~(\ref{s}) and then, if
necessary, solve the bulk equations in the neighborhood of the brane with the
given boundary data on the brane.

One way of doing this consists in choosing the Gaussian normal coordinates in
the bulk generated by the Minkowski coordinates $x^\alpha$, $\alpha =
0,\ldots,3$, on the brane, so that the metric of the solution in these
coordinates has the form
\begin{equation}
d s^2_5 = \epsilon dy^2 + h_{\alpha\beta} (y, x) dx^\alpha dx^\beta \, ,
\end{equation}
where $y \ge 0$ is the fifth coordinate in the bulk, and the brane corresponds
to $y = 0$. Introducing also the tensor of extrinsic curvature with components
$K_{\alpha\beta}$ on every hypersurface $y = {\rm const}$ in the bulk, one can
obtain the following system of differential equations for the components
$h_{\alpha\beta}$ and $K^\alpha{}_\beta$:
\begin{eqnarray} \label{curv}
{\partial K^\alpha{}_\beta \over \partial y} &=& \epsilon \left[
R^\alpha{}_\beta - \frac16 \left(R + 2 \Lambda \right) \delta^\alpha{}_\beta
\right] - K K^\alpha{}_\beta - \frac16 \left( K^\mu{}_\nu K^\nu{}_\mu - K^2
\right) \delta^\alpha{}_\beta \nonumber \\ &=& \epsilon \left( R^\alpha{}_\beta
- \frac23 \Lambda \delta^\alpha{}_\beta \right) - K K^\alpha{}_\beta \, ,
\end{eqnarray}
\begin{equation} \label{metric}
{\partial h_{\alpha\beta} \over \partial y} = 2 h_{\alpha\gamma}
K^\gamma{}_\beta \, ,
\end{equation}
where $R^\alpha{}_\beta$ are the components of the Ricci tensor of the metric
$h_{\alpha\beta}$ induced on the hypersurface $y = {\rm const}$, $R =
R^\alpha{}_\alpha$ is its scalar curvature, and $K = K^\alpha{}_\alpha$ is the
trace of the tensor of extrinsic curvature. The second equality in (\ref{curv})
is true by virtue of the constraint equation (\ref{constraint}).  Equations
(\ref{curv}) and (\ref{metric}) together with the constraint equation
(\ref{constraint}) represent the $4\!+\!1$ splitting of the Einstein equations
in the Gaussian normal coordinates.  Solution of Eqs.~(\ref{curv}) and
(\ref{metric}) in the bulk always exists in some neighborhood of the brane.

We illustrate this method on the example of the generalization of the
Randall--Sundrum (RS) solution \cite{RS} to the case of arbitrary signature of
the fifth dimension and arbitrary Ricci-flat brane.  In this solution, the
matter on the brane is in the vacuum state, and the induced metric on the brane
is Ricci-flat, so that $\tau_{\alpha\beta} = 0$ and $R^\alpha{}_\beta = 0$ at
the brane. It follows from Eq.~(\ref{brane}) that $K^\alpha{}_\beta \propto
\delta^\alpha{}_\beta$ at the brane.  Then Eqs.~(\ref{curv}) and (\ref{metric})
imply that $K^\alpha{}_\beta (y, x) \equiv \frac14 K (y) \delta^\alpha{}_\beta$
and the induced metric $h_{\alpha\beta} (y, x)$ is Ricci-flat on every
hypersurface $y = {\rm const}$, so that $R^\alpha{}_\beta (y, x) \equiv 0$.
After this, Eqs.~(\ref{constraint}) and (\ref{brane}) give
\begin{equation} \label{rs}
K^2 = - \frac83 \epsilon \Lambda\, , \quad K = - {4 \epsilon \sigma \over 3
M^3}  \quad \Longrightarrow \quad \epsilon \Lambda = - {2 \sigma^2 \over 3 M^6}
\, ,
\end{equation}
which, in particular, means that $\epsilon \Lambda$ must be negative. Now
Eq.~(\ref{metric}) can easily be solved, and the resulting metric is
\begin{equation}
ds^2 = \epsilon dy^2 + \exp \left(- {\epsilon \sigma \over 3 M^3}\, y \right)
h_{\alpha\beta} (x) dx^\alpha dx^\beta \, ,
\end{equation}
where $h_{\alpha\beta} (x)$ are the components of the Ricci-flat metric on the
brane.

If one is interested in the evolution on the brane only, and if no additional
boundary and regularity conditions are specified for the bulk, then it suffices
to solve only Eq.~(\ref{closed}) together with the matter equations on the
brane. As an example, let us show that the cosmological equations for the brane
follow immediately from Eq.~(\ref{closed}).  We consider a homogeneous and
isotropic cosmological model with the cosmological time $t$, scale factor $a
(t)$, energy density $\rho (t)$, and stress $p (t)$.  First, using the
stress--energy conservation equation, we obtain
\begin{equation}
M^6 a^3 \dot a \left(S_{ab} S^{ab} - \frac13 S^2\right) = - \frac13 {d \over
dt} \left[ 3 m^2 \left(\dot a^2 + \kappa \right) - a^2 (\rho + \sigma)
\right]^2 \, ,
\end{equation}
where $\kappa = 0, \pm 1$ indicates the spatial curvature of the model. The
expression $a^3 \dot a (R - 2 \Lambda)$ is also a total derivative; hence,
Eq.~(\ref{closed}) can be integrated once with the result (obtained by
different methods in \cite{CH,Deffayet} for the case of $\epsilon = 1$)
\begin{equation} \label{cosmo}
m^4 \left( H^2 + {\kappa \over a^2} - {\rho + \sigma \over 3 m^2} \right)^2 =
\epsilon M^6  \left(H^2 + {\kappa \over a^2} - {\Lambda \over 6} - {C \over
a^4} \right) \, ,
\end{equation}
where  $C$ is the integration constant and $H \equiv \dot a/a$.  For $m = 0$,
this equation passes to a generalization of the well-known result
\cite{BDL,BDEL}:
\begin{equation}\label{cosmolim}
H^2 + {\kappa \over a^2} = {\Lambda \over 6} + {C \over a^4} + \epsilon {(\rho
+ \sigma)^2 \over 9 M^6}\, .
\end{equation}
Some consequences of this equation for $\epsilon = -1$ are discussed in
\cite{SS}. For $M = 0$, Eq.~(\ref{cosmo}) becomes the standard cosmological
equation in Einstein's theory.

Thus far, we considered the case where a brane is the boundary of a
five-dimensional manifold.  Now we will obtain a closed system of equations for
a brane which is a junction between two volume spaces with the cosmological
constants $\Lambda_1$ and $\Lambda_2$, respectively, and without imposing $Z_2$
symmetry. Brane-world cosmologies without $Z_2$ symmetry of reflection were a
subject of thorough study, for example, in \cite{nonsym1,nonsym2,CH,Kofinas}.
Our study will be slightly more complete in the sense that we retain the
curvature term in the action on the brane, i.e., in our case, $m \ne 0$. The
braneworld model without $Z_2$ symmetry was studied in \cite{KLM} in the case
of $m \ne 0$ and $\Lambda_1 = \Lambda_2$.

Equation (\ref{brane}) now is modified as follows:
\begin{equation} \label{junction}
m^2 G_{ab} + \sigma h_{ab} = \epsilon M^3 \left(S^{(1)}_{ab} + S^{(2)}_{ab}
\right) + \tau_{ab} \, ,
\end{equation}
where the tensors $S^{(1)}_{ab}$ and $S^{(2)}_{ab}$ are expressed through
the extrinsic curvatures of the brane in two volumes, respectively. (We
choose the normals $n_1^a$ and $n_2^a$ with respect to which we calculate
the extrinsic curvatures to be the inner normals in the corresponding
volumes.) We introduce the sum $S_{ab} = S^{(1)}_{ab} + S^{(2)}_{ab}$ so
that Eq.~(\ref{junction}) has the same form as Eq.~(\ref{brane}).

Equation (\ref{constraint}) is now satisfied by $S^{(i)}_{ab}$ in the place
of $S_{ab}$ and $\Lambda_i$ in the place of $\Lambda$ for $i = 1,2$.
Introducing also the difference $Q_{ab} = S^{(1)}_{ab} - S^{(2)}_{ab}$, we
easily obtain the following closed system of gravitational equations on the
brane:
\begin{equation} \label{constraint-new}
\epsilon \left(R - \Lambda_1 - \Lambda_2 \right) + \frac14 \left(S_{ab} S^{ab}
- \frac13 S^2 \right) + \frac14 \left(Q_{ab} Q^{ab} - \frac13 Q^2 \right) = 0
\, ,
\end{equation} \begin{equation} \label{orthog}
S_{ab} Q^{ab} - \frac13 S Q = 2 \epsilon \left( \Lambda_1 - \Lambda_2 \right)
\, ,
\end{equation} \begin{equation} \label{conserve-new}
D_a Q^a{}_b = 0 \, ,
\end{equation}
where $Q = Q_{ab} h^{ab}$, and $S_{ab}$ is formally given by the same equation
as (\ref{s}):
\begin{equation} \label{s-new}
S_{ab} = {\epsilon \over M^3} \left(  m^2 G_{ab} + \sigma h_{ab} - \tau_{ab}
\right) \, .
\end{equation}
Equation (\ref{conserve-new}) follows from relation (\ref{conserve-s}). This
system of equations is to be solved for the metric and matter fields and for
the symmetric tensor field $Q_{ab}$ on the brane. Imposing $Z_2$ symmetry
corresponds to setting $Q_{ab} = 0$ and $\Lambda_1 = \Lambda_2$ and is
equivalent to the case of brane as a boundary (edge) considered above and
described by Eq.~(\ref{closed}) (these two cases are related by rescaling
the constant $M^3$ by a factor of two).

We consider briefly the cosmological implications of system
(\ref{constraint-new})--(\ref{s-new}).  Of course, even with the homogeneous
and isotropic cosmological setting for the ordinary matter and gravity, the
symmetric tensor $Q_{ab}$, in general, need not have homogeneous and
isotropic form. However, here, for simplicity, we set it to be homogeneous
and isotropic so that
\begin{equation}
Q^0{}_0 = \beta (t) \, , \qquad Q^\mu{}_\nu = \delta^\mu{}_\nu q (t) \, ,
\quad \mu, \nu = 1,2,3 \, .
\end{equation}
Then Eqs.~(\ref{orthog}) and (\ref{conserve-new}) easily yield the condition
\begin{equation} \label{alpha}
2 m^2 \beta \left( H^2 + {\kappa \over a^2}  - {\rho + \sigma \over 3 m^2}
\right) = M^3 \left( \Lambda_1 - \Lambda_2 + {E \over a^4} \right) \, ,
\end{equation}
where $E$ is the integration constant.  Now, using the conservation equation
(\ref{conserve-new}), we have
\begin{equation}
a^3 \dot a \left( Q_{ab}Q^{ab} - \frac13 Q^2 \right) = - \frac13 {d \over
dt} \left(a^4 \beta^2 \right) \, .
\end{equation}
Finally, a procedure quite similar to that which led us to Eq.~(\ref{cosmo})
gives a partial generalization of Eq.~(\ref{cosmo}) to the case where $Z_2$
symmetry is not imposed:
\begin{eqnarray}
m^4 \left( H^2 + {\kappa \over a^2} - {\rho + \sigma \over 3 m^2} \right)^2 &=&
4 \epsilon  M^6  \left(H^2 + {\kappa \over a^2} - {\Lambda_1 + \Lambda_2 \over
12} - {C \over a^4} \right)  \nonumber \\ &-& {M^{12} \over 36 m^4} \left[
{\Lambda_1 - \Lambda_2 + E/a^4 \over H^2 + \kappa / a^2 - (\rho + \sigma) /
3m^2 } \right]^2 ,
\end{eqnarray}
where we used Eq.~(\ref{alpha}) to eliminate $\beta$. A similar equation was
derived in \cite{KLM} by the standard method of embedding for the partial case
$\epsilon = 1$ and $\Lambda_1 = \Lambda_2$. The case of $Z_2$ symmetry
corresponds here to setting $\Lambda_1 = \Lambda_2$ and $E = 0$. For $m = 0$,
this equation reduces to a generalization of (\ref{cosmolim}),
\begin{equation}
H^2 + {\kappa \over a^2} = {\Lambda_1 + \Lambda_2 \over 12} + {C \over a^4} +
\epsilon {(\rho + \sigma)^2 \over 36 M^6} + \epsilon {M^6 \over 16}
\left({\Lambda_1 - \Lambda_2 + E / a^4 \over \rho + \sigma} \right)^2 \, ,
\end{equation}
which in various forms appeared in \cite{nonsym1} for the case of $\epsilon =
1$.

We conclude this work.  The system of equations
(\ref{constraint-new})--(\ref{s-new}) is a closed system that describes the
evolution on a brane.  This system is all one has to solve in a theory in which
there are no restrictions on solutions in the five-dimensional bulk. If there
{\em are\/} such restrictions in the form of boundary and/or regularity
conditions, then system (\ref{constraint-new})--(\ref{s-new}) cannot be
regarded as complete: only those its solutions are admissible which can be
developed to solutions in the bulk satisfying the imposed regularity
conditions. This illustrates the importance of the precise specification of
boundary and/or regularity conditions in the bulk for a complete formulation of
the braneworld theory: in the absence of such specification, {\em all\/} the
information about the brane is contained in the closed system
(\ref{constraint-new})--(\ref{s-new}).

Finally, we note that system (\ref{constraint-new})--(\ref{s-new}) admits
solutions with flat vacuum branes without fine tuning the constants of the
theory even in the case $\Lambda_1 = \Lambda_2$, although it requires the
absence of $Z_2$ orbifold symmetry and, as a consequence, absence of the
Lorentz symmetry of the bulk \cite{Shtanov2}.

This work was supported in part by the INTAS grant for project No.~2000-334.

\end{document}